\documentstyle[12pt]{article}

\begin{document}
\begin{titlepage}
\begin{flushleft}
Stockholm\\
USITP 95\\
June 1995\\
\end{flushleft}
\vspace{1cm}
\begin{center}
{\Large NEIGHBOURS OF EINSTEIN'S EQUATIONS:}\\
\ \\
{\Large CONNECTIONS AND CURVATURES}\\
\vspace{2cm}
{\large Ingemar Bengtsson}\footnote{Email address: ingemar@vana.physto.se}\\
{\sl Fysikum\\
University of Stockholm\\
Box 6730, S-113 85 Stockholm, Sweden}\\
\vspace{3cm}
{\bf Abstract}\\
\ \\
\end{center}

\

\noindent Once the action for Einstein's equations is rewritten as a
functional of an SO(3, {\bf C}) connection and a conformal factor of
the metric, it admits a family of ``neighbours'' having the same number
of degrees of freedom and a precisely defined metric tensor. This paper
analyzes the relation between the Riemann tensor of that metric and the
curvature tensor of the SO(3) connection. The relation is in general
very complicated. The Einstein case is distinguished by the fact that
two natural SO(3) metrics on the GL(3) fibers coincide. In the general
case the theory is bimetric on the fibers.  \\

\end{titlepage}

\noindent {\bf 1. INTRODUCTION.}

\vspace*{5mm}

\noindent The principles of general relativity include requirements such
 as diffeomorpism invariance and a dynamically determined space-time
metric. It is of philosophical interest to ask to what extent the form of
 Einstein's equations are enforced by these principles. We want to know
how tightly they constrain the world. It is also well to keep in mind a
more practical reason for trying to modify Einstein's equations; indeed
there is observational evidence (notably galactic rotation curves) which,
 if taken at face value, suggests that the equations fail at large
distances. Of course, the most likely explanation of this evidence is
that we have a ``dark matter'' problem on our hands, rather than a
problem with Einstein's equations, but nevertheless these observations
do add further interest to the question of the degree of uniqueness of
the latter.

In 1915, Einstein was under the impression that he had found the unique
field equations embodying the principles of general relativity. This
first impression has stood the test of time rather well. Although a one
 parameter ambiguity was found early on (with the parameter christened
 ``the cosmological constant''), most later attempts at generalization
 have involved rather drastic changes in the theory. At the very least,
 new degrees of freedom are added to the theory, as in scalar-tensor
theories and theories based on higher derivative actions. The latter
have some manifestly unphysical features. A more radical, and as yet
unfinished, construction is provided by string theory. This involves a
 very large number of additional degrees of freedom, and presumably
also some changes in the underlying principles - as should be true for
all ``unified'' theories. The only attempt known to the author in which
 Einstein's equations are generalized with no additional degrees of
freedom - except for the well known family of theories parametrized by
 the cosmological constant - is a class of possible generalizations of
 Einstein's equations to which we refer as ``neighbours of Einstein's
equations''. The field equations can be derived from an action which is
 a functional of an SO(3,{\bf C}) connection $A_{{\alpha}i}$ and a
scalar field ${\eta}$ with tensor density weight minus one, and the
action is of the form \cite{CDJ} \cite{Physlett}

\begin{equation} S[A,{\eta}] =
 \int {\cal L}({\eta},Tr{\bf {\Omega}},Tr{\bf {\Omega}}^2,
Tr{\bf {\Omega}}^3) \ ,
 \label{1}  \end{equation}

\noindent where the function ${\cal L}$ is chosen so that the integrand
 has density weight one, ``$Tr$'' denotes an SO(3) trace, and

\begin{eqnarray} F_{{\alpha}{\beta}}^{\ \ i} &=&
 \partial_{\alpha}A_{\beta}^{\ i} - {\partial}_{\beta}A_{\alpha}^{\ i} +
 i{\gamma}^{im}{\epsilon}_{mjk}A_{\alpha}^{\ j}A_{\beta}^{\ k} \\
\ \nonumber \\
{\Omega}^{ij} &=& \frac{1}{2}{\epsilon}^{{\alpha}{\beta}{\gamma}{\delta}}
F_{{\alpha}{\beta}}^{\ \ i}F_{{\gamma}{\delta}}^{\ \ j} \ . \end{eqnarray}

\noindent Here the inverse of the SO(3) metric ${\gamma}_{ij}$ occurs
explicitly. We are being pedantic about this since we will later
introduce an {\it a priori} different fibre metric $m_{ij}$. Our
conventions for raising and lowering indices will be carefully spelled
 out in section 2. Except for pathological choices of the function
${\cal L}$, it is easy to cast this action in Hamiltonian form
\cite{C} \cite{B&P} \cite{Physlett}. It will then be seen that for
generic choices of ${\cal L}$ there are two degrees of freedom per
space-time point. Moreover, since the space-time metric can be
constructed from the structure functions of the constraint algebra, one
 can calculate a metric tensor which, when rewritten in manifestly
four-dimensional notation, is

\begin{equation} g_{{\alpha}{\beta}} =
 \frac{4}{3}{\eta}{\epsilon}_{ijk}{\epsilon}^{{\mu}{\nu}{\rho}{\sigma}}
F_{{\alpha}{\mu}}^{\ \ i}F_{{\nu}{\rho}}^{\ \ j}F_{{\sigma}{\beta}}^{\ \ k}
 . \label{4} \end{equation}

\noindent This formula holds whenever the field equations hold, and the
expression is unique up to conformal transformations, so that the
conformal structure is uniquely determined in a space-time which solves
the field equations. We stress that the equation is not a definition of
the metric tensor, it is the statement of a theorem \cite{B&P}
\cite{JMP}. A further theorem then follows immediately, namely that
the SO(3) field strengths are self-dual with respect to this metric
\cite{Schonberg}.

It is known that the original Einstein equations are equivalent to the
field equations following from the above action, provided that
${\cal L}$ is suitably specified \cite{CDJ}:

\begin{equation} S[A, {\eta}] = \frac{1}{2} \int
\frac{\eta}{\sqrt{\gamma}}(Tr{\bf {\Omega}}^2
- \frac{1}{2}(Tr{\bf {\Omega}})^2)
\ . \end{equation}

\noindent (Actually, this is not quite true, since the formalism breaks
 down for Petrov types $\{3,1\}$, $\{4\}$ and $\{-\}$, but this is the
 least of our problems, and we will ignore this point in the sequel.)
Another highly intricate choice of ${\cal L}$ gives equations
equivalent to Einstein's with a non-vanishing cosmological constant
 \cite{Peldan}. The Einstein case is distinguished in that the two
fibre metrics (${\gamma}_{ij}$ and $m_{ij}$) coincide, whereas in
general they do not. We will return to this point in section 5.

What we have just described is very much an unfinished construction, for
several reasons:

1) While we know how to characterize solutions with real Lorentzian
metrics, and some exact solutions with a real Lorentzian metric are
known, we have no useful characterization of the set of such solutions,
which makes comparison to the special Einstein case difficult.

2) It is not known whether propagation is causal with respect to the
metric that we have identified.

3) It is technically difficult to add matter degrees of freedom to the
theory, and only some very special cases have been worked out.

The construction will remain an unfinished one also at the end of the
present paper. The mathematical problem that we will solve here is how
 to relate the Riemann tensor of the metric to the SO(3) field strength
 that occurs in the action. Actually, ``study'' may be more
appropriate than ``solve'', since no computationally useful formula
will emerge. In the Einstein case, the SO(3) field strength is simply
 the self dual part of the Riemann tensor, but in the general case the
 relation is decidedly complicated, as we will see. The problem will
 be formulated more precisely in section 2, where some further
background information can be found as well. In section 3 we extend
the SO(3) covariant derivative to a derivative which is covariant under
 local GL(3) transformations of the fibres as well as under space-time
 diffeomorphisms. We also demand that the extended derivative shall be
 ``compatible'' with a triad of two-forms and a scalar density. The
formalism that we use is based on a peculiarly four dimensional
variation of Riemannian geometry which uses triads of two-forms rather
 than tetrads of vector fields, and which is due to 'tHooft
\cite{'tHooft} \cite{IB}. (See also ref. \cite{Lunev}.) In section 4 it
 will be shown how to use this formalism to establish a relation
between the metric Riemann tensor and the SO(3) curvature. The
establishment of this formula was one of the main goals of the
investigation that is reported here - naturally I hoped that a fairly
 simple formula would emerge, but this hope was crossed by the facts.
 In section 5 we discuss the dynamical equations derived for a one
parameter family of neighbours of Einstein's equations; in particular
 we will see how the two fibre metrics are related to each other. Then
 we give some examples of exact solutions of the same equations:
``Kasner-like'' solutions in section 6 and ``Schwarzschild-like'' in
section 7. In the concluding section 8 we summarize the argument (for
 the benefit of those readers who do not want to lose themselves among
 calculational details) and make some further comments.

\vspace*{1cm}

\noindent {\bf 2. NEIGHBOURS AND NOTATION.}

\vspace*{5mm}

\noindent Here we will develop the ``neighbours of Einstein's
equations'' a little bit further, in order to make our notation clear.
 There will be not only an SO(3) fiber over every space-time point,
but a fiber equipped with two different metrics ${\gamma}_{ij}$ and
$m_{ij}$, which define two different SO(3) subspaces of GL(3). This
feature will put a strain on our notation.

First we make clear that ``$Tr$'' in eq. (\ref{1}) means

\begin{equation} Tr{\bf {\Omega}} \equiv {\gamma}_{ij}{\Omega}^{ij} \ ,
 \end{equation}

\noindent and we stress that ${\gamma}_{ij}$ is some metric that we are
allowed to specify in any way we wish (in most cases, one would set it
equal to Kronecker's delta).

Before we derive the field equations from the action (\ref{1}), we make
two definitions:

\begin{eqnarray} {\Psi}_{ij} &\equiv &
 \frac{\partial {\cal L}}{\partial {\Omega}^{ij}} \label{7} \\
\ \nonumber \\
{\Sigma}_{{\alpha}{\beta}i} &\equiv &
 {\Psi}_{ij}F_{{\alpha}{\beta}}^{\ \ j} \ . \label{8} \end{eqnarray}

\noindent Then the field equations take the form

\begin{eqnarray} {\epsilon}^{{\alpha}{\beta}{\gamma}{\delta}}
D_{\beta}{\Sigma}_{{\gamma}{\delta}i} &=& 0 \label{9} \\
\ \nonumber \\
\frac{\partial {\cal L}}{\partial {\eta}} &=& 0 \label{10} \ .
\end{eqnarray}

\noindent (The second equation here is equivalent to the Hamiltonian
constraint in the canonical formulation.)

We will analyze these equations further later on. Eq. (\ref{9}), which is
 shared by all the neighbours (and also turns up in other contexts), will
 be dealt with in the following two sections, while eqs. (\ref{8}) and
 (\ref{10}) will be discussed for a special one parameter family of
neighbours in section 5. For the moment, we use the triad ${\Sigma}_i$
 of two-forms to introduce a second metric on the fibers, namely

\begin{equation} m_{ij} \equiv
\frac{1}{2}{\epsilon}^{{\alpha}{\beta}{\gamma}{\delta}}
{\Sigma}_{{\alpha}{\beta}i}{\Sigma}_{{\gamma}{\delta}j} \ . \end{equation}

\noindent We will now adopt the convention that we use ${\gamma}_{ij}$
to raise and lower the latin indices on the objects

\begin{equation} A_{\alpha}^{\ i} \hspace*{15mm} F_{{\alpha}{\beta}}^{\ \ i}
\hspace*{15mm} {\Omega}^{ij} \ , \end{equation}

\noindent and $m_{ij}$ to raise and lower the Latin indices on all other
 objects except the ${\epsilon}$-tensors. Greek indices will be raised
 and lowered with the space-time metric, which is unique. The inverse
 and determinant of ${\gamma}_{ij}$ are denoted respectively by
${\gamma}^{ij}$ and ${\gamma}$, and similarly for the other metrics.
 An over-riding convention is that all our ${\epsilon}$-tensors take
the values $\pm 1$ in every coordinate system, and hence their indices
are never raised or lowered with any metric. A further useful piece of
 notation is

\begin{equation} \tilde{\Sigma}^{{\alpha}{\beta}} \equiv
\frac{1}{2}{\epsilon}^{{\alpha}{\beta}{\gamma}{\delta}}
{\Sigma}_{{\gamma}{\delta}} \ . \end{equation}

As shown in refs. \cite{B&P} \cite{JMP}, in any solution of the
equations the space-time metric is

\begin{equation} g_{{\alpha}{\beta}} =
 \frac{8}{3}{\eta}{\epsilon}_{ijk}F_{{\alpha}{\gamma}}^{\ \ i}
\tilde{F}^{{\gamma}{\delta}j}F_{{\delta}{\beta}}^{\ \ k} =
\frac{8}{3}{\sigma}{\epsilon}^{ijk}{\Sigma}_{{\alpha}{\gamma}i}
\tilde{\Sigma}^{{\gamma}{\delta}}_{\ \ j}
{\Sigma}_{{\delta}{\beta}k}  \ , \end{equation}

\noindent where

\begin{equation} {\sigma} = \frac{\eta}{\det{\Psi}} \ .
\label {15} \end{equation}

\noindent This metric has a very special geometrical significance
\cite{Schonberg}: It can be defined (uniquely up to a conformal factor)
 by the requirement that the $F_i$'s, as well as the ${\Sigma}_i$'s,
span the subspace of self-dual two-forms. Or stated the other way
around, in any solution of the equations these objects are self-dual
 two-forms. For our purposes it is of course essential to know under
 which conditions on the (complex valued) two-forms the metric is real
 and Lorentzian. This condition turns out to be

\begin{equation} {\epsilon}^{{\alpha}{\beta}{\gamma}{\delta}}
F_{{\alpha}{\beta}}^{\ \ i}\bar{F}_{{\gamma}{\delta}}^{\ \ j} = 0 \ ,
\label{17} \end{equation}

\noindent where the bar denotes complex conjugation. An extra condition
 is needed to ensure the reality of the conformal factor.

The basic idea is that the ${\Sigma}_i$'s, or alternatively the $F_i$'s,
 by construction form a basis for the three dimensional space of
self-dual two-forms. (The CDJ formalism then fails for certain
algebraically degenerate field configurations, for which the $F_i$'s
 are not linearly independent.) I reviewed these matters recently
\cite{IB}\footnote{I have to add that the preprint version of ref.
\cite{IB} contains an embarrassing sign error.}, and do not propose
 to do so again here, but two useful formul\ae \ must be recorded:

\begin{equation} \tilde{\Sigma}_{{\alpha}{\beta}} =
 4{\sigma}^2m{\Sigma}_{{\alpha}{\beta}} \end{equation}

\begin{equation} {\Sigma}_{{\alpha}{\gamma}i}
\tilde{\Sigma}^{\gamma}_{\ {\beta}j} = - \frac{1}{4}m_{ij}
g_{{\alpha}{\beta}}
 - \frac{1}{4{\sigma}}{\epsilon}_{ijk}
\tilde{\Sigma}_{{\alpha}{\beta}}^{\ \ k} \ .
\end{equation}

\noindent These formul\ae \ were used liberally in the calculations to
 be reported below.

\vspace*{1cm}

\noindent {\bf 3. COVARIANT DERIVATIVES.}

\vspace*{5mm}

\noindent Let us recall the starting point, which is that we have
available an SO(3) covariant derivative defined by

\begin{equation} D_{\alpha}V_i = \partial_{\alpha}V_i + i{\epsilon}_{ijk}
A_{\alpha}^{\ j}{\gamma}^{km}V_m \ . \end{equation}

\noindent Note the presence of the imaginary unit; actually we are
dealing with an SO(3,{\bf C}) connection here. The group metric is
denoted by ${\gamma}_{ij}$; we do not use it to raise and lower
indices because we will shortly introduce another metric which we
will use for such a purpose.

It is assumed that

\begin{equation} D_{[{\alpha}}{\Sigma}_{{\beta}{\gamma}]i} = 0 \ .
\end{equation}

\noindent And the problem is to use this equation to set up a relation
between the curvature tensor $F_{{\alpha}{\beta}i}$ and the Riemann
tensor of the metric

\begin{equation}
g_{{\alpha}{\beta}} = \frac{8}{3}{\sigma}{\epsilon}^{ijk}
{\Sigma}_{{\alpha}{\gamma}i}\tilde{\Sigma}^{{\gamma}{\delta}}_{\ \ j}
{\Sigma}_{{\delta}{\beta}k} \ . \end{equation}

\noindent Note that no special relation between $F_{{\alpha}{\beta}i}$
 and ${\Sigma}_{{\alpha}{\beta}i}$ is assumed in this or in the
following section. Conversely, any special relation is allowed.

To solve the problem, it will be convenient to introduce a number of
 new covariant derivatives, which extend the SO(3) covariant derivative
 which is presented to us at the outset. First we extend $D$ to a GL(3)
 covariant derivative

\begin{equation} {\cal D}_{\alpha}V_{{\beta}i} =
D_{\alpha}V_{{\beta}i} + {\beta}_{{\alpha}i}^{\ \ j}V_{{\beta}j} \ .
 \end{equation}

\noindent Then we introduce no less than three different GL(4) covariant
 derivatives, viz.

\begin{equation} \nabla^{(g)}_{\alpha}V_{{\beta}i} =
 \partial_{\alpha}V_{{\beta}i} -
{\scriptsize \left\{ \begin{array}{c} {\gamma} \\
{\alpha}{\beta} \end{array} \right\}} V_{{\gamma}i}  \end{equation}

\noindent (this makes use of the Christoffel symbol, i.e. of an affine
 connection that is a function of the metric),

\begin{equation} \nabla^{({\Gamma})}_{\alpha}V_{{\beta}i} =
\partial_{\alpha}V_{{\beta}i} - {\Gamma}_{{\alpha}{\beta}}^{\ \ {\gamma}}
V_{{\gamma}i} \end{equation}

\noindent (defined using a symmetric affine connection), and

\begin{equation} \nabla^{(T)}_{\alpha}V_{{\beta}i} =
\nabla^{({\Gamma})}_{\alpha}V_{{\beta}i} +
\frac{1}{2}T_{{\alpha}{\beta}}^{\ \ {\gamma}}V_{{\gamma}i} \ ,
\end{equation}

\noindent where the torsion tensor is self-dual, which means that it
 is of the form

\begin{equation} T_{{\alpha}{\beta}}^{\ \ {\gamma}} =
{\Sigma}_{{\alpha}{\beta}i}T^{{\gamma}i} \ . \end{equation}

\noindent Finally, we extend $D$ to a derivative which is both GL(3)
 and GL(4) covariant:

\begin{equation} \nabla_{\alpha}V_{{\beta}i} = D_{\alpha}V_{{\beta}i}
 + {\beta}_{{\alpha}i}^{\ \ j}V_{{\beta}j} -
 {\Gamma}_{{\alpha}{\beta}}^{\ \ {\gamma}}V_{{\gamma}i} +
 \frac{1}{2}T_{{\alpha}{\beta}}^{\ \ {\gamma}}V_{{\gamma}i} \ .
 \end{equation}

The idea is that the various connections that we have introduced -
 altogether $12 + 24 + 40 + 12 = 88$ unknown components - will be
 determined through the $12 + 72 + 4 = 88$ conditions

\begin{eqnarray} D_{[{\alpha}}{\Sigma}_{{\beta}{\gamma}]i} = 0
\hspace*{55mm}
\label{30} \\
\ \nonumber \\
\ \nabla_{\alpha}{\Sigma}_{{\beta}{\gamma}i} =
D_{\alpha}{\Sigma}_{{\beta}{\gamma}i} +
 {\beta}_{{\alpha}i}^{\ \ j}{\Sigma}_{{\beta}{\gamma}j} -
 {\Gamma}_{{\alpha}{\beta}}^{\ \ {\delta}}{\Sigma}_{{\delta}{\gamma}i} -
 {\Gamma}_{{\alpha}{\gamma}}^{\ \ {\delta}}{\Sigma}_{{\beta}{\delta}i} +
 \frac{1}{2} T_{{\alpha}{\beta}}^{\ \ {\delta}}{\Sigma}_{{\delta}{\gamma}i}
 + \frac{1}{2} T_{{\alpha}{\gamma}}^{\ \ {\delta}}
{\Sigma}_{{\beta}{\delta}i} = 0
\label{31} \\
\ \nonumber \\
\nabla_{\alpha}{\sigma} = \partial_{\alpha}{\sigma}
 - {\beta}_{{\alpha}i}^{\ \ i}{\sigma} +
 {\Gamma}_{{\alpha}{\gamma}}^{\ \ {\gamma}}{\sigma} -
 \frac{1}{2}T_{{\alpha}{\gamma}}^{\ \ {\gamma}}{\sigma} = 0 \hspace*{3cm}
\label{32} \end{eqnarray}

\noindent The status of these conditions is the same as that of the
``metric postulate'' in the customary presentation of Riemannian
geometry. The task at hand is to solve these equations for the
connections that we have introduced. This is easily done in a step by
step analysis.

We make the definition

\begin{equation} t_{{\alpha}i} \equiv
 {\gamma}{\epsilon}_{ijk}{\gamma}^{km}(\partial_{\alpha}{\Sigma}_
{{\beta}{\gamma}m} + \partial_{\gamma}{\Sigma}_{{\alpha}{\beta}m} +
\partial_{\beta}{\Sigma}_{{\gamma}{\alpha}m})
\tilde{\Sigma}^{{\beta}{\gamma}j} \ . \end{equation}

\noindent Then we have

\begin{eqnarray} D_{[{\alpha}}{\Sigma}_{{\beta}{\gamma}]i} &=& 0 \\
\Leftrightarrow \nonumber \\
A_{{\alpha}i}({\Sigma}) &=& \frac{3i}{2}t_{{\alpha}i} +
\frac{i}{2{\sigma}m}m_{im}{\epsilon}^{mjk}
\tilde{\Sigma}_{{\alpha}\ j}^{\ {\beta}}t_{{\beta}k}\\
\ \nonumber \\
\ \nonumber \\
\nabla_{\alpha}({\sigma}m_{ij}) &=& 0 \\
\Leftrightarrow \nonumber \\
{\beta}_{{\alpha}i}^{\ \ j}({\sigma},{\Sigma})
 &=& \frac{1}{2{\sigma}}({\delta}_i^jm^{mn}D_{\alpha}({\sigma}m_{mn}) -
 m^{jk}D_{\alpha}({\sigma}m_{ik})) \\
\ \nonumber \\
\ \nonumber \\
\nabla_{[{\alpha}}{\Sigma}_{{\beta}{\gamma}]i} &=& 0 \\
\Leftrightarrow \nonumber \\
T^{{\alpha}i}({\sigma},{\Sigma}) &=&
 2(\tilde{\Sigma}^{{\alpha}{\beta}i}{\beta}_{{\beta}j}^{\ \ j} -
\tilde{\Sigma}^{{\alpha}{\beta}j}{\beta}_{{\beta}j}^{\ \ i}) \\
\ \nonumber \\
\ \nonumber \\
\nabla_{\alpha}g_{{\beta}{\gamma}} &=& 0 \\
\Leftrightarrow \nonumber \\
{\Gamma}_{{\alpha}{\beta}}^{\ \ {\gamma}}({\sigma}, {\Sigma}) &=&
{\scriptsize \left\{ \begin{array}{c} {\gamma} \\
{\alpha}{\beta} \end{array} \right\}}
- \frac{1}{2}T^{\gamma}_{\ \ {\alpha}{\beta}} -
\frac{1}{2}T^{\gamma}_{\ \ {\beta}{\alpha}} \ , \end{eqnarray}

\noindent where the Christoffel symbol is

\begin{equation} {\scriptsize \left\{ \begin{array}{c} {\gamma} \\
{\alpha}{\beta} \end{array} \right\}} =
\frac{1}{2}g^{{\gamma}{\delta}}(\partial_{\beta}g_{{\delta}{\alpha}} +
 \partial_{\alpha}g_{{\delta}{\beta}} -
 \partial_{\delta}g_{{\alpha}{\beta}}) \ .
 \end{equation}

\noindent This solves the problem at hand, which was to ensure that the
connections are ``compatible'' with ${\Sigma}$ and ${\sigma}$ in the
sense of eqs. (\ref{31})-(\ref{32}), while at the same time they extend
a given SO(3) connection obeying eq. (\ref{30}).

In refs. \cite{'tHooft} and \cite{IB} similar ideas were applied to
formulate Einstein's equations. That is a simpler task than the one
which occupies us here, because then the GL(4) connection can be chosen
to be torsion free.

\vspace*{1cm} {\bf 4. CURVATURE TENSORS.}

\vspace*{5mm}

\noindent For each of the covariant derivatives that were introduced in
the previous section there is a curvature tensor. First there is an
SO(3) curvature

\begin{equation} [D_{\alpha}, D_{\beta}]V_{{\gamma}i} =
 i{\epsilon}_{ijk}F_{{\alpha}{\beta}}^{\ \ j}{\gamma}^{km}V_{{\gamma}m}
 \end{equation}

\noindent and a GL(3) curvature

\begin{eqnarray} [{\cal D}_{\alpha}, {\cal D}_{\beta}]V_{{\gamma}i} =
 {\cal F}_{{\alpha}{\beta}i}^{\ \ \ j}V_{{\gamma}j} = \hspace*{7cm}
\nonumber \\
\ \\
= (i{\epsilon}_{imn}F_{{\alpha}{\beta}}^{\ \ m}{\gamma}^{nj} +
D_{\alpha}{\beta}_{{\beta}i}^{\ \ j} -
D_{\beta}{\beta}_{{\alpha}i}^{\ \ j} +
[{\beta}_{\alpha}, {\beta}_{\beta}]_i^{\ j})V_{{\gamma}j} \ .
\nonumber \end{eqnarray}

\noindent Then the three different affine connections that we introduced
 each give rise to a Riemann tensor, as follows:

\begin{eqnarray} [\nabla^{(g)}_{\alpha}, \nabla^{(g)}_{\beta}]
V_{{\gamma}i} &=&
 R^{\ \ \ \ {\delta}}_{{\alpha}{\beta}{\gamma}}V_{{\delta}i} \\
\ \nonumber \\
{[}\nabla^{({\Gamma})}_{\alpha}, \nabla^{({\Gamma})}_{\beta}]V_{{\gamma}i}
&=& R^{({\Gamma}) \ {\delta}}_{{\alpha}{\beta}{\gamma}}V_{{\delta}i} \\
\ \nonumber \\
{[}\nabla^{(T)}_{\alpha}, \nabla^{(T)}_{\beta}]V_{{\gamma}i} &=&
 R^{(T) \ {\delta}}_{{\alpha}{\beta}{\gamma}}V_{{\delta}i} +
T_{{\alpha}{\beta}}^{\ \ {\delta}}\nabla_{\delta}V_{{\gamma}i} \ .
\end{eqnarray}

\noindent Finally

\begin{equation} [\nabla_{\alpha}, \nabla_{\beta}]V_{{\gamma}i} =
 R^{(T) \ {\delta}}_{{\alpha}{\beta}{\gamma}}V_{{\delta}i} +
T_{{\alpha}{\beta}}^{\ \ {\delta}}\nabla_{\delta}V_{{\gamma}i}
 + {\cal F}_{{\alpha}{\beta}i}^{\ \ \ j}V_{{\gamma}j} \ .
\end{equation}

\noindent (I hope that the reader symphatizes with the way I have tried
 to solve the notational difficulties here.)

We will also define the self-dual parts of the various Riemann tensors
according to

\begin{equation} R_{{\alpha}{\beta}i} \equiv
R_{{\alpha}{\beta}{\gamma}{\delta}}\tilde{\Sigma}^{{\gamma}{\delta}}_{\ \ i}
\ , \end{equation}

\noindent and similarly for the other cases. Given
$R_{{\alpha}{\beta}i}$, the full metric Riemann tensor can be
reconstructed by means of complex conjugation, provided that
condition (\ref{17}) for real Lorentzian metrics is fullfilled.
For the other Riemann tensors this is not true, since (for instance)

\begin{equation} R^{(T)}_{{\alpha}{\beta}{\gamma}{\delta}} \neq
 - R^{(T)}_{{\alpha}{\beta}{\delta}{\gamma}} \end{equation}

\noindent in general. However, our aim is to set up a relation between
$F_{{\alpha}{\beta}i}$ and $R_{{\alpha}{\beta}i}$, and therefore
this need not concern us.

We proceed with the calculation; we use the equation

\begin{equation}
[\nabla_{\alpha}, \nabla_{\beta}]{\Sigma}_{{\gamma}{\delta}i} =
R^{(T) \ {\sigma}}_{{\alpha}{\beta}{\gamma}}
{\Sigma}_{{\sigma}{\delta}i}
 - R^{(T) \ {\sigma}}_{{\alpha}{\beta}{\delta}}
{\Sigma}_{{\sigma}{\gamma}i}
 + {\cal F}_{{\alpha}{\beta}i}^{\ \ \ j}{\Sigma}_{{\gamma}{\delta}j} = 0
  \end{equation}

\noindent to relate the self-dual part of the Riemann tensor, for that
connection which includes torsion, to the GL(3) curvature tensor. The
result is

\begin{eqnarray} {\cal F}_{{\alpha}{\beta}ij} &=&
- \frac{1}{2}R^{(T)\ {\gamma}}_{{\alpha}{\beta}{\gamma}}m_{ij}
+ \frac{1}{2{\sigma}}{\epsilon}_{ijk}R^{(T)k}_{{\alpha}{\beta}} \\
\ \nonumber \\
R^{(T)i}_{{\alpha}{\beta}} &=&
{\sigma}{\epsilon}^{ijk}{\cal F}_{{\alpha}{\beta}jk}
 \ . \label{51} \end{eqnarray}

It is now straightforward but tedious to extract the desired relation.
We will quote some intermediate steps for reference; round and square
brackets denote symmetrization and anti-symmetrization, respectively,
both with weight one. Thus:

\begin{eqnarray} R_{{\alpha}{\beta}{\gamma}}^{(T) \ {\delta}} &=&
 R_{{\alpha}{\beta}{\gamma}}^{({\Gamma}) \ {\delta}} +
\nabla_{[{\alpha}}T_{{\beta}]{\gamma}}^{\ \ \ {\delta}} -
\frac{1}{2}T_{{\sigma}[{\alpha}}^{\ \ \ {\delta}}
T_{{\beta}]{\gamma}}^{\ \ \ {\sigma}} -
\frac{1}{2}T_{{\alpha}{\beta}}^{\ \ {\sigma}}
T_{{\sigma}{\gamma}}^{\ \ {\delta}} \\
\ \nonumber \\
\ \nonumber \\
R_{{\alpha}{\beta}{\gamma}}^{({\Gamma}) \ {\delta}} &=&
 R_{{\alpha}{\beta}{\gamma}}^{\ \ \ \ {\delta}} +
2\nabla_{[{\alpha}}T^{\delta}_{\ ({\beta}]{\gamma})}
 - T_{{\alpha}{\beta}}^{\ \ {\sigma}}T^{\delta}_{\ ({\gamma}{\sigma})}
 + T_{{\gamma}[{\alpha}}^{\ \ \ {\sigma}}T^{\delta}_{\ ({\beta}]{\sigma})} -
\nonumber \\
\ \\
&\ & \hspace*{35mm}  - T_{{\sigma}[{\alpha}}^{\ \ \ {\delta}}
T^{\sigma}_{\ ({\beta}]{\gamma})}
+ 2T^{\delta}_{\ ({\sigma}[{\alpha})}T^{\sigma}_{\ ({\beta}]{\gamma})}
\nonumber \\
\ \nonumber \\
R^{(T)}_{{\alpha}{\beta}i} &=& R_{{\alpha}{\beta}i} + \frac{2}{\sigma}
\tilde{\Sigma}_{[{\alpha}}^{\ \ {\gamma}j}
\nabla_{{\beta}]}D_{\gamma}({\sigma}m_{ij}) + \nonumber \\
\ \nonumber \\
&+& \frac{1}{4}{\Sigma}_{[{\alpha}}^{\ \ {\gamma}j}T_{{\beta}]j}
T_{{\gamma}i}
 + \frac{1}{4{\sigma}}{\epsilon}_{ijk}{\Sigma}_{{\gamma}
[{\alpha}}^{\ \ \ m}\tilde{\Sigma}_{{\beta}]{\delta}}^{\ \ \ j}
T^{{\gamma}k}T^{\delta}_{\ m} + \nonumber \\
\ \label{54} \\
&+& \frac{1}{2}{\Sigma}_{{\gamma}[{\alpha}i}T_{{\beta}]j}T^{{\gamma}j}
+ \frac{1}{4{\sigma}}{\epsilon}_{ijk}T_{\alpha}^{\ j}T_{\beta}^{\ k}
+ \nonumber \\
\ \nonumber \\
&+& \frac{1}{8}{\Sigma}_{{\alpha}{\beta}}^{\ \ j}
(7T_{{\gamma}j}T^{\gamma}_{\ i} + \frac{2}{\sigma}{\epsilon}_{imn}
T_{{\gamma}j}\tilde{\Sigma}^{{\gamma}{\delta}m}T_{\delta}^{\ n}
+ \frac{1}{\sigma}{\epsilon}_{jmn}T_{\gamma}^{\ m}
\tilde{\Sigma}^{{\gamma}{\delta}}_{\ \ i}T_{\delta}^{\ n})
- \frac{1}{8}{\Sigma}_{{\alpha}{\beta}i}T_{\gamma}^{\ j}T^{\gamma}_{\ j}
 \ . \nonumber \end{eqnarray}

\noindent Unfortunately, although we have manipulated this expression
further, we have been unable to simplify it much. Therefore the best
summary we can give at this point is to say that an explicit formula
for the Riemann tensor as a function of the ${\Sigma}_i$'s and
${\sigma}$ can be obtained by using eqs. (\ref{51}) together with
(\ref{54}). This solves our problem (in principle).

\vspace*{1cm}

\noindent {\bf 5. THE DYNAMICAL EQUATIONS.}

\vspace*{5mm}

\noindent The problem that was discussed in the two previous sections
is a geometrical one, with no dynamical restrictions on the two-forms
imposed. The precise form of the action (\ref{1}) enters when we turn
to the dynamical field equations, that is to say when we assume that

\begin{equation} {\Sigma}_{{\alpha}{\beta}i} =
{\Psi}_{ij}F_{{\alpha}{\beta}}^{\ \ j} \ , \end{equation}

\noindent with the matrix ${\Psi}_{ij}$ given by eq. (\ref{7}). The field
equation (\ref{9}), together with the Bianchi identity for the field
strength, can be rewritten in the form

\begin{equation} \tilde{F}^{{\alpha}{\beta}j}D_{\beta}{\Psi}_{ij} = 0 \
. \end{equation}

\noindent We also add the constraint (\ref{10}) that ensures that the
action is stationary with respect to variations in the field ${\eta}$.
It is out of the question to discuss the general case here, so we will
concentrate on a one parameter family of neighbours of Einstein's
equations given by the action \cite{C} \cite{B&P}

\begin{equation} S[A, {\eta}] =
\frac{1}{2} \int \frac{\eta}{\sqrt{\gamma}}(Tr{\bf {\Omega}}^2 +
{\alpha}(Tr{\bf {\Omega}})^2) \ . \end{equation}

\noindent Note that the parameter ${\alpha}$ is dimensionless, and that
${\alpha} = - 1/2$ gives Einstein's vacuum equations \cite{CDJ}. As far
as I know, the results that follow are typical - at least I did not
choose this special action with any deliberate intention of obtaining
especially simple results. Anyway, we will not go very far into the
subject - essentially only one conclusion will be drawn.

To begin, we obtain from the definition in eq. (\ref{7}) that

\begin{equation} {\Psi}_{ij} =
\frac{\eta}{\sqrt{\gamma}}({\Omega}_{ij}
+ {\alpha}{\gamma}_{ij}Tr{\bf {\Omega}})
 \ . \end{equation}

\noindent (Recall that the fixed fibre metric ${\gamma}_{ij}$ is used
to raise and lower indices on ${\Omega}_{ij}$.) The consraint
(\ref{10}) becomes

\begin{equation} Tr{\bf {\Omega}}^2 + {\alpha}(Tr{\bf {\Omega}})^2 = 0 \ .
\label{61} \end{equation}

\noindent It will also be useful to recall the characteristic equation
for three-by-three matrices:

\begin{equation} {\bf  M}^3 - {\bf M}^2Tr{\bf M}
- \frac{1}{2}{\bf M}(Tr{\bf M}^2 - (Tr{\bf M})^2) - \det{\bf M} =
0 \ . \end{equation}

Now we are in a position to express the fibre metric $m_{ij}$ in terms
of the SO(3) field strengths. Using the characteristic equation in
conjunction with the constraint (\ref{61}), we find that

\begin{eqnarray} m_{ij} &\equiv &
{\Sigma}_{{\alpha}{\beta}i}\tilde{\Sigma}^{{\alpha}{\beta}}_{\ \ j} =
{\Psi}_{im}{\Omega}^{mn}{\Psi}_{nj} = \nonumber \\
\ \nonumber \\
&=& \frac{{\eta}^2}{\gamma}({\gamma}_{ij}\det{\bf {\Omega}} +
({\alpha} + \frac{1}{2})({\alpha} - 1)(Tr{\bf {\Omega}})^2{\Omega}_{ij} +
(1 + 2{\alpha})Tr{\bf {\Omega}}{\Omega}^2_{ij}) = \\
\ \nonumber \\
&=& {\sigma}^2m({\gamma}_{ij} + (1 + 2{\alpha})\frac{Tr{\bf {\Omega}}}
{\det {\bf {\Omega}}}({\Omega}_{ij}^2 + \frac{1}{2}({\alpha} - 1)
Tr{\bf {\Omega}}{\Omega}_{ij})) .
\end{eqnarray}

If we choose ${\alpha} = - 1/2$ this equation simplifies dramatically,
and the two fibre metrics are then proportional. It is easy to
show that

\begin{equation} {\alpha} = - \frac{1}{2} \hspace*{1cm} \Rightarrow
\hspace*{1cm} {\sigma}m_{ij} = \frac{1}{\sqrt{\gamma}} {\gamma}_{ij} \ .
\end{equation}

\noindent Now the calculations in section 4 collapse:

\begin{equation} {\alpha} = - \frac{1}{2} \hspace*{1cm} \Rightarrow
\hspace*{1cm} {\beta}_{{\alpha}i}^{\ \ j} = 0 \hspace*{1cm} \Rightarrow
\hspace*{1cm} R_{{\alpha}{\beta}}^{\ \ i} =
- \frac{2i}{\sqrt{\gamma}}F_{{\alpha}{\beta}}^{\ \ i} \ . \end{equation}

\noindent So we see that the Einstein case (${\alpha} = - \frac{1}{2}$)
is indeed very special, in that when the field equations hold the two
{\it a priori} different metrics that were introduced on the fibers
coincide. This brings further simplifications in train; thus the SO(3)
field strength that occurs in the action equals the self-dual part of
the Riemann tensor of the spacetime metric. In turn, since the SO(3)
field strengths span the space of self-dual two-forms, this implies that
the self-dual part of the Riemann tensor is self-dual also in its
remaining part of indices, which is an exceptional situation which holds
if and only if the traceless part of the Ricci tensor vanishes.

In the general case we have a theory that is bimetric on the fibres,
even though there is a unique conformal structure in space-time, as
determined by the Hamiltonian constraint algebra \cite{B&P}. To our
disappointment we have been unable to use the specific choice of the
action to simplify eq. (\ref{54}) for the Riemann tensor in any
significant way.

\vspace*{1cm}

      \noindent {\bf 6. KASNER-LIKE SOLUTIONS.}

\vspace*{5mm}

\noindent To add concreteness to the above, we will consider some exact
solutions of the equations that were considered in the previous section.
 Throughout, we make the convenient choice that

\begin{equation} {\gamma}_{ij} = {\delta}_{ij} \ . \end{equation}

To obtain our first example, we make an Ansatz which is known to give the
familiar Kasner solution in the Einstein case, namely \cite{CP}

\begin{equation} A_{x1} = a_1(t) \hspace*{1cm} A_{y2} = a_2(t)
\hspace*{1cm} A_{z3} = a_3(t) \ , \end{equation}

\noindent all other components vanishing. After some manipulation, one
finds that the field equations collapse to

\begin{equation} \left(\frac{\dot{a}_1}{a_1}\right)^2
+ \left(\frac{\dot{a}_2}{a_2}\right)^2 +
\left(\frac{\dot{a}_3}{a_3}\right)^2 +
{\alpha}\left(\frac{\dot{a}_1}{a_1} + \frac{\dot{a}_2}{a_2} +
\frac{\dot{a}_3}{a_3}\right)^2 = 0 \end{equation}

\begin{equation} \partial_t({\eta}a_2a_3\dot{a}_1) =
\partial_t({\eta}a_3a_1\dot{a}_2) = \partial_t({\eta}a_1a_2\dot{a}_3)
= 0 \ . \end{equation}

\noindent A convenient choice of gauge is

\begin{equation} {\eta} = - \frac{1}{16\dot{a}_1\dot{a}_2\dot{a}_3}
\ . \end{equation}

\noindent Then the solution is

\begin{equation}
a_1 = d_1t^{{\gamma}_1 - 1} \hspace*{1cm}
a_2 = d_2t^{{\gamma}_2 - 1} \hspace*{1cm}
a_3 = d_3t^{{\gamma}_3 - 1} \hspace*{1cm}
\ , \end{equation}

\noindent where the ``Kasner exponents'' obey

\begin{equation} {\gamma}_1 + {\gamma}_2 + {\gamma}_3 = 1 \hspace*{15mm}
{\gamma}_1^2 + {\gamma}_2^2 + {\gamma}_3^2 = - 1 - 4{\alpha}
\  \end{equation}

\noindent We choose the integration constants so that we obtain a simple
expression for the space-time metric - which means that certain other
expressions appear to be more complicated than they are, because one
can not have everything. Specifically, we choose

\begin{equation}
d_1 = \sqrt{({\gamma}_2 - 1)({\gamma}_3 - 1)} \hspace*{8mm}
d_2 = \sqrt{({\gamma}_3 - 1)({\gamma}_1 - 1)} \hspace*{8mm}
d_3 = \sqrt{({\gamma}_1 - 1)({\gamma}_2 - 1)} \ . \end{equation}

\noindent This ensures that the spacetime metric takes the familiar
Kasner form

\begin{equation} ds^2 = - dt^2 + t^{2{\gamma}_1}dx^2 +
t^{2{\gamma}_2}dy^2 + t^{2{\gamma}_3}dz^2 \ . \end{equation}

\noindent For our other geometrical objects, we obtain

\begin{equation}
{\Sigma}_{tx1} = \frac{i}{4d_1}(1 + 2{\alpha} - {\gamma}_1)t^{{\gamma}_1}
 \hspace*{15mm} {\Sigma}_{yz1} = \frac{1}{4d_1}({\gamma}_1 - 1 - 2{\alpha})
t^{1 - {\gamma}_1} \ , \end{equation}

\noindent and similarly for ${\Sigma}_{ty2}, {\Sigma}_{tz3}$ and
${\Sigma}_{zx2}, {\Sigma}_{xy3}$ (and so on below). Further,

\begin{equation} {\sigma} =
4i\frac{({\gamma}_1 - 1)({\gamma}_2 - 1)({\gamma}_3 - 1)}
{({\gamma}_1 - 1 - 2{\alpha})({\gamma}_2 - 1 - 2{\alpha})
({\gamma}_3 - 1 - 2{\alpha})}t^{-1} \hspace*{3cm} \end{equation}

\begin{equation} {\sigma}m_{ij} = {\delta}_{ij} - 2(1 + 2{\alpha})\left(
\begin{array}{ccc} \frac{{\gamma}_1 + {\alpha}}{({\gamma}_2 - 1 - 2{\alpha})
({\gamma}_3 - 1 - 2{\alpha})} & 0 & 0 \\
0 & \frac{{\gamma}_2 + {\alpha}}{({\gamma}_3 - 1 - 2{\alpha})
({\gamma}_1 - 1 - 2{\alpha})} & 0 \\
0 & 0 & \frac{{\gamma}_3 + {\alpha}}{({\gamma}_1 - 1 - 2{\alpha})
({\gamma}_2 - 1 - 2{\alpha})}\end{array}\right) \end{equation}

\begin{equation}
R_{tx1} = \frac{1}{2d_1}{\gamma}_1(1 - {\gamma}_1)
({\gamma}_1 - 1 - 2{\alpha})t^{{\gamma}_1 - 1} \hspace*{12mm} R_{yz1}
= \frac{i}{2d_1}{\gamma}_2{\gamma}_3(1 + 2{\alpha} - {\gamma}_1)
t^{- {\gamma}_1} \ . \end{equation}

\noindent We observe that there is only one non-vanishing component
of the Ricci tensor, viz.

\begin{equation} R_{tt} = 2(1 + 2{\alpha})t^{-2} \ . \end{equation}

\vspace*{1cm}

\noindent {\bf 7. SCHWARZSCHILD-LIKE SOLUTIONS.}

\vspace*{5mm}

\noindent We will give one more explicit example of a geometry, namely
static and spherically symmetric solutions. The derivation was given in
ref. \cite{B&P}, and here we will confine ourselves to stating the
result. In order to make a spherically symmetric Ansatz, it is
convenient to use spherical polar coordinates, and to introduce the vectors

\begin{eqnarray} U_i = (\sin{\theta}\cos{\phi}, \sin{\theta}\sin{\phi},
 \cos{\theta}) \hspace*{1cm} V_i = (\cos{\theta}\cos{\phi}, \cos{\theta}
\sin{\phi}, - \sin{\theta}) \nonumber \\
\ \\
W_i = ( - \sin{\phi}, \cos{\phi}, 0) \ . \hspace* {45mm} \nonumber
\end{eqnarray}

\noindent A form of the solution which is the most general one compatible
with a spherically symmetric and static line element is then given by

\begin{equation} \begin{array}{lll} A_{ti} = - \frac{(2M)^{1-c}}{\sqrt{2c}}
(1 - {\varphi}^2)^cU_i
& \hspace*{1cm} & A_{ri} = 0 \\
\ \\
A_{{\theta}i} =
i(1 - {\varphi})W_i & \hspace*{1cm} & A_{{\phi}i} =
- i\sin{\theta}(1 - {\varphi})V_i \end{array} \end{equation}

\begin{equation} {\eta} = \frac{\sqrt{2c}(2M)^{2c - 1}}{16}
\frac{(1 - {\varphi}^2)^{-2c - 1}}{{\varphi}{\varphi}'\sin{\theta}}
\hspace*{5mm} \end{equation}

\noindent where ${\varphi} = {\varphi}(r)$ is an arbitrary function of
$r$, the slash denotes differentiation with respect to $r$, $M$ is an
integration constant (adjusted to simplify the metric), and

\begin{equation} c = \frac{1}{1 + {\alpha}}(\pm \sqrt{- \frac{1}{2}
(1 + 3{\alpha})} - {\alpha}) \ . \end{equation}

\noindent The fact that there are two roots of this equation may appear
to contradict Birkhoff's theorem on the uniqueness of the Schwarzschild
solution, but appearances deceive. In the Einstein case,
(${\alpha} = - 1/2$), what happens is that $c = 2$ gives the
Schwarzschild solution while $c = 0$ gives a degenerate metric. In the
generic case both roots give rise to non-degenerate metrics, but then
there is no Birkhoff's theorem to evade.

The results are

\begin{eqnarray}
F_i &\equiv & \frac{1}{2}dx^{\alpha}dx^{\beta}F_{{\alpha}{\beta}i}
= - \sqrt{2c}(2M)^{1 - c}{\varphi}{\varphi}'(1 - {\varphi}^2)^{c-1}dtdrU_i
 + \nonumber \\
\ \nonumber \\
&+& \frac{(2M)^{1-c}}{\sqrt{2c}}{\varphi}(1 - {\varphi}^2)^c
(dtd{\theta}V_i + \sin{\theta}dtd{\phi}W_i) - \\
\ \nonumber \\
&-& i{\varphi}'(drd{\theta}W_i + \sin{\theta}d{\phi}drV_i) +
i(1 - {\varphi}^2)\sin{\theta}d{\theta}d{\phi}U_i \nonumber \\
\ \nonumber \\
\ \nonumber \\
{\Sigma}_i &=&
i\sqrt{2c}M(c +{\alpha}c + {\alpha}){\varphi}{\varphi}'
(1 - {\varphi}^2)^{-2}dtdrU_i - \nonumber \\
\ \nonumber \\
&-& \frac{iM}{2\sqrt{2c}}(2{\alpha}c + 1 + 2{\alpha}){\varphi}
(1 - {\varphi}^2){-1}(dtd{\theta}V_i + \sin{\theta}dtd{\phi}W_i) -
\nonumber \\
\ \\
&-& \frac{(2M)^c}{4}(2{\alpha}c + 1 + 2{\alpha}){\varphi}'
(1 - {\varphi}^2)^{-c-1}(drd{\theta}W_i + \sin{\theta}d{\phi}drV_i) +
\nonumber \\
\ \nonumber \\
&+& \frac{(2M)^c}{2}(c + {\alpha}c + {\alpha})(1 - {\varphi}^2)^{-c}
\sin{\theta}d{\theta}d{\phi}U_i \nonumber \\
\ \nonumber \\
\ \nonumber \\
{\sigma} &=& - \frac{4i(2M)^{-c-1}}
{\sqrt{2c}(1 + 3{\alpha})(2{\alpha}c + 1 + 2{\alpha})}
\frac{(1 - {\varphi}^2)^{c+2}}{{\varphi}{\varphi}'\sin{\theta}} \\
\ \nonumber \\
\ \nonumber \\
ds^2 &=& - (2M)^{2-c}{\varphi}^2(1 - {\varphi}^2)^{c-2}
dt^2 + \nonumber \\
\ \\
&+& 2c(2M)^c({\varphi}')^2(1 - {\varphi}^2)^{-c-2}dr^2 +
(2M)^c(1 - {\varphi}^2)^{-c}(d{\theta}^2 + \sin^2{\theta}d{\phi}^2)
 \nonumber \\
\ \nonumber \\
\ \nonumber \\
{\sigma}m_{ij} &=& \frac{1}{2(c + {\alpha}c + {\alpha})}
({\delta}_{ij} - (4c + 6{\alpha}c + 1 + 6{\alpha})U_iU_j) \\
\ \nonumber \\
\ \nonumber \\
R_{{\alpha}{\beta}i} &=& 4(2M)^2(c + 2{\alpha}c + \frac{1}{2} +
2{\alpha}){\varphi}^2({\varphi}')^2
\frac{(1 - \frac{c}{2}{\varphi}^2)}{(1 - {\varphi}^2)^{4}}
\sin{\theta}dtdrU_i - \nonumber \\
\ \nonumber \\
&-& \frac{(2M)^2}{4}(2{\alpha}c + 1 + 2{\alpha}){\varphi}^2{\varphi}'
\frac{(1 + (1 - c){\varphi}^2)}{(1 - {\varphi}^2)^{3}}\sin{\theta}
(dtd{\theta}V_i + \sin{\theta}dtd{\phi}W_i) + \nonumber \\
\ \\
&+& \frac{i}{4}\sqrt{2c}(2M)^{c+1}(2{\alpha}c + 1 + 2{\alpha})
\frac{{\varphi}({\varphi}')^2}{(1 - {\varphi}^2)^{c+2}}\sin{\theta}
(drd{\theta}W_i + \sin{\theta}d{\phi}drV_i) + \nonumber \\
\ \nonumber \\
&+& i\sqrt{2c}(2M)^{1+c}(c + {\alpha}c + {\alpha})
{\varphi}{\varphi}'\frac{(1 - \frac{c}{2}{\varphi}^2)}
{(1 - {\varphi}^2)^{2+c}}d{\theta}d{\phi}U_i \ . \nonumber
\end{eqnarray}

\noindent The only non-vanishing components of the Ricci tensor are

\begin{equation}
R_{tt} = (\frac{2}{c} - 1)(2M)^{2 - 2c}{\varphi}^2(1 - {\varphi}^2)^{2c - 2}
 \hspace*{1cm}
R_{rr} = 2(c - 2)(2M)({\varphi}')^2\frac{(1 - c{\varphi}^2)}
{(1 - {\varphi}^2)^2}
 \ . \end{equation}

\noindent Finally, the ``Schwarzschild'' form of the line element is
obtained if we choose the free function ${\varphi}(r)$ according to

\begin{equation} {\varphi}^2 = 1 - \frac{2M}{r^{2/c}} \ . \end{equation}

\vspace*{1cm}

\noindent {\bf 8. CONCLUSIONS.}

\vspace*{5mm}

\noindent Let us make a brief sketch of the argument. The action
(\ref{1}), with a general choice of the integrand ${\cal L}$, gives
field equations of the form

\begin{equation} D_{[{\alpha}}{\Sigma}_{{\beta}{\gamma}]i} = 0
\end{equation}

\begin{equation} \frac{\partial {\cal L}}{\partial {\eta}} = 0 \ ,
\end{equation}

\noindent where the ${\Sigma}_i$'s are definite functions of ${\eta}$
and the $F_i$'s. From an analysis of the constraint algebra in the
Hamiltonian formulation, we conclude that the space-time metric is
given by the Sch\"{o}nberg-Urbantke expression, eq. (\ref{4}), which
 guarantees that the two-forms are self-dual. We found in this paper
 that the SO(3) covariant derivative may be extended to a derivative
 which is covariant under both local GL(3) transformations of the
fibers as well as space-time diffeomorphisms, and that the extension
 becomes unique when we require that this derivative annihilates both
 the triad ${\Sigma}_i$ of two-forms and the conformal factor
${\sigma}$ of the metric. It was also shown that one can use these
compatibility conditions to express the Riemann tensor of this
derivative, as well as the metric Riemann tensor, as a definite function
 of the two-forms, the conformal factor, and the SO(3) curvature tensor.
 The expression is unwieldy, however.

We then concentrated our attention to a special one parameter family of
 action integrands, which includes the special choice which gives rise
 to Ricci flat metrics, and analyzed the relation between the two
natural metrics on the fibers, the inert metric ${\gamma}_{ij}$ which
 is part of the definition of the theory, and the dynamical metric
$m_{ij}$ that is defined by the triad of two-forms. It was found that
 they coincide in the Einstein case, and only in the Einstein case.
Finally some exact solutions of the equations were examined.

Although no computationally useful formula for the Riemann tensor
emerged in the non-Einstein case, the results do give insight into
the structure of ``neighbour geometry''.

It is legitimate to ask why we regard $m_{ij}$ rather than
${\Omega}_{ij}$ as a natural fibre metric. The answer is that in the
Einstein case the latter matrix is degenerate for certain field
configurations, such as flat Minkowski space, for which the field
strengths $F_i$'s fail to provide a basis for the space of self-dual
 two-forms. On the other hand the ${\Sigma}_i$'s do form such a basis
 by assumption, since again the space-time metric is non-degenerate
by assumption. To this one may object that this is an unnatural
restriction within the present framework, and moreover that the
situation may be different in the non-Einstein case. I do not have
 a good retort to this objection available. It is of course possible
 to replace ${\Sigma}_i$ and ${\sigma}$ by $F_i$ and ${\eta}$ thoughout
 the calculations of sections 3 and 4, but as far as I can see this
 leads to no significant simplifications.

How does our construction evade Lovelock's theorem, which is a strong
 uniqueness theorem for Einstein's equations in four space-time
dimensions \cite{Love}? Suppose that we rewrite our ``vacuum''
equations as an equation which has the Einstein tensor of the metric
 on the left hand side, and something else on the right hand side.
Without going into details, we remark - for the benefit of those who
more or less remember the theorem - that apart from the fact that the
 right hand side is not directly given as a functional of the metric,
 we also escape Lovelock's theorem because the divergence of the right
 hand side is zero only as a consequence of the field equations.
Indeed the metric takes the Sch\"{o}nberg-Urbantke form only when the
 equations of motion hold. It is therefore understandable that only
partial results \cite{B&P} \cite{new results} are known for matter
couplings.

Finally, the phenomenological prospect for the neighbours is known to
 be bleak \cite{B&P} \cite{Ola}. However, if I am allowed to end on a
 speculative note, this prospect may change in unpredictable ways if
 one is able to turn the fibre metric ${\gamma}_{ij}$ in the action
into a dynamical field. Which seems a natural thing to try, anyway.

\end{document}